\documentclass[10pt,letterpaper]{article}
\usepackage{opex3}
\usepackage{cite}

\usepackage{ae} 

\begin{document}

\title{Steady-periodic method for modeling mode instability in fiber amplifiers}

\author{Arlee V. Smith$^*$ and Jesse J. Smith}

\address{AS-Photonics, LLC, 8500 Menaul Blvd. NE, Suite B335, Albuquerque, NM 87112 USA}

\email{$^*$arlee.smith@as-photonics.com}

\begin{abstract}
We present a detailed description of the methods used in our model of mode instability in high-power, rare earth-doped, large-mode-area fiber amplifiers. Our model assumes steady-periodic behavior, so it is appropriate to operation after turn on transients have dissipated. It can be adapted to transient cases as well. We describe our algorithm, which includes propagation of the signal field by fast-Fourier transforms, steady-state solutions of the laser gain equations, and two methods of solving the time-dependent heat equation: alternating-direction-implicit integration, and the Green's function method for steady-periodic heating.
\end{abstract}


\section{Introduction}

In earlier papers we described a physical processes that can lead to modal instability\cite{smithsmith2011} in high power fiber amplifiers. We also presented simulation results that illustrated how periodic modulation of the pump or signal seed can dramatically reduce the instability threshold\cite{smithsmith2012a}. Although we described our model and methods in those papers we did not present the mathematical details. Here we do so. The individual parts of the model are all known, but the way they are combined to form a steady-periodic model of modal instability is original.

First, we offer a brief reprise of the physics included in our fiber amplifier model. Mode instability is the degradation of output beam quality above a sharp power threshold\cite{Otto}. Below threshold the light emerges in the fundamental mode, above threshold some is in higher order modes, usually LP$_{11}$. Assuming most of the signal seed light is injected into the fundamental mode LP$_{01}$ with a small amount populating a higher order mode, usually mode LP$_{11}$, these two modes interfere along the length of the fiber. Because they have different propagation constants their interference creates a signal irradiance pattern that oscillates along the length of the fiber. Pump light is preferentially absorbed in regions of higher signal irradiance, and because a fraction of the absorbed pump is converted to heat, this creates a heating pattern that resembles the irradiance pattern. The heat pattern is converted to a similar temperature pattern, and the temperature pattern creates a refractive index pattern via the thermo optic effect. If the interference pattern is stationary, there is little or no phase shift between the thermally-induced index pattern and the irradiance pattern, so there is almost zero net transfer of power between modes. However, if the light in the higher order mode is slightly detuned in frequency from that in LP$_{01}$, the irradiance pattern moves along the fiber - down stream for a red detuning and up stream for a blue detuning. The temperature pattern also moves, but it lags the irradiance pattern, and this lag produces the phase shift necessary for power transfer between modes. Red detuning of LP$_{11}$ leads to power transfer from LP$_{01}$ to LP$_{11}$. The detuning that maximizes the mode coupling is set largely by the thermal diffusion time across the fiber core. This is approximately 1 ms, implying an optimum frequency detuning of approximately 1 kHz.

When light in LP$_{01}$ is transferred to LP$_{11}$ by deflection from the moving grating it experiences a frequency shift equal to the frequency offset, due to a Doppler effect\cite{smithsmith2011}, so the transferred light adds coherently to the light already in LP$_{11}$. This mechanism can cause a substantial transfer of the power from LP$_{01}$ to LP$_{11}$ if the pump power exceeds a sharp mode instability threshold. This gain process can be categorized as near-forward stimulated thermal Rayleigh scattering.

Our modeling approach is to develop as general a model as feasible, and our model includes all the physical effects just described. We use diffractive beam propagation of a steady-periodic, time-dependent signal field. This field incorporates all fiber modes (including lossy modes) and all offset frequencies that are harmonics of the primary frequency offset. The time-dependent thermal profile is computed using either an alternating-direction-implicit (ADI) integration method or a steady-periodic Green's function method. Mode coupling occurs through inclusion of the thermally-induced refractive index change in the index profile that is used to compute beam propagation. No analytic or semi-analytic expressions of mode coupling are needed, and coupling between all modes is included. We use this approach because it permits the most accurate modeling, and because it makes adding various physical effects relatively straightforward. The cost of such a general numerical model is long run-times, but using the methods described here one can model several meters of fiber per hour on a consumer grade desktop computer.

Alternative mode coupling models based on similar physical effects have been published by other authors\cite{Hansen,Jauregui,Ward} but we will not review them here.

\section{The scalar, paraxial beam propagation equation}

Equation (\ref{eqn:paraxial_wave_eqn}) is the scalar, paraxial beam propagation equation for an isotropic medium. It is derived in numerous papers on nonlinear fiber optics, for example \cite{Feit1980_1,Feit1980_2}. The variable $E_s$ is a complex envelope function that represents all spatial and temporal modulation of a monochromatic, plane carrier wave. The direction of propagation is $z$. This equation is appropriate for small index contrasts typical of step index, large mode area fiber amplifiers. 
\begin{equation}\label{eq.parax}
\label{eqn:paraxial_wave_eqn} \frac{\partial E_s(x,y,z,t)}{\partial z} = \frac{i}{2 k_c}\nabla^2_\perp E_s(x,y,z,t) + \frac{i\big[ k^2(x,y,z,t)-k_c^2 \big]}{2k_c}E_s(x,y,z,t)
\end{equation}
where $k_c=\omega_c n_{\rm clad}/c$ is the wave number of the carrier wave in the cladding, and $k(x,y,z,t)=\omega_c n(x,y,z,t)/c$. Here $n(x,y,z,t)$ is the guiding index including the thermo-optic contribution,
\begin{equation}\label{eqn:core_profile}
n(x,y,z,t)=n_{\rm core}(x,y,z,t) + \frac{{\rm d}n}{{\rm d}T} T(x,y,z,t),
\end{equation}
where ${\rm d}n/{\rm d}T$ is the thermo optic coefficient given in Table~\ref{tab:input_params}. The first term on the right hand side of Eq.~(\ref{eq.parax}) describes diffraction in a homogeneous medium with a refractive index equal to the cladding index $n_c$. Here $\nabla_{\perp}^2$ is the Laplacian operator in the transverse dimensions $x$ and $y$. The second term adds a correction to account for the core guidance, including the usual core refractive index step plus the thermally induced index change. In general $k(x,y,z,t)$ is imaginary to account for gain or loss, but we will use $k(x,y,z,t)$ to represent only the real part, and write the much smaller imaginary part as a separate term with the coefficient $g(x,y,z,t)$, 
\begin{equation}
\label{eqn:paraxial_wave_eqn_with_gain} \frac{\partial E(x,y,z,t)}{\partial z} = \frac{i}{2 k_c}\nabla^2_\perp E(x,y,z,t) + \frac{i\big[ k^2(x,y,z,t)-k_c^2  \big]}{2k_c}E(x,y,z,t) + g(x,y,z,t) E(x,y,z,t).
\end{equation}

\section{Integrating the propagation equation}

Equation (\ref{eqn:paraxial_wave_eqn_with_gain}) can be integrated by various methods - for example, using the split step method described below, or using finite-differences. Whichever method is used, the central issue is how to compute the thermally induced part of $k(x,y,z,t)$ for use in the $z$ integration.

One approach is to integrate the field along the full length of the fiber at $t=0$ based on an initial temperature profile $T(x,y,z,0)$. The temperature profile is then updated based on the heat deposited during the time increment $\Delta t$, plus the previous temperature profile and the thermal boundary conditions. The new temperature profile is used to integrate the field at time $\Delta t$ along the full length of the fiber. This iteration of full $z$ integration of $E_s$ alternating with stepping $T$ by $\Delta t$ is repeated for successive time steps to find the temperature and field histories, both over the domain $(x,y,z,t)$. 

Alternatively, a time sequence of fields at $z=0$ can be used to compute $T(x,y,t)$ at the input and that time varying temperature can be used to propagate the time varying field by $\Delta z$. This is repeated for successive $z$ steps. One limitation is that the temperature is solved in two dimensions rather than three so longitudinal heat flow is not accounted for.

\subsection{Steady-periodic condition}

Both of the methods just described can treat transient or steady-periodic cases. However, our goal is to compute behavior only under steady-periodic conditions. We are not interested in starting from an initial transient and following the amplifier evolution to a steady state since that may require integration over a time longer than the thermal diffusion time from the core to the outer diameter of the fiber. The required settling time can be greater than 100 ms, perhaps much greater, depending on the fiber design.

Because transient effects are of secondary interest, and also in order to achieve shorter run times, we base our model on the assumption that all transients have decayed and the only time dependence is periodic modulation of powers, mode content, temperature, etc. With this steady-periodic assumption we need integrate only over a single modulation period of approximately 1 ms. We first perform the time integration on the temperature equation to find $T(x,y,t)$ for a single period at one $z$ position, then we step the signal and pump fields by $\Delta z$.

The steady-periodic condition appears to correspond well with reported behavior of amplifiers operating near and slightly above their mode instability thresholds. Far above threshold the mode coupling may be more complicated\cite{Otto}. The relevant time scale for the steady-periodic condition is the time for heat diffusion over the core rather than over the entire fiber. The temperature outside the core is determined by the time averaged heating in the core, but it cannot respond rapidly enough to follow the periodic heating variations within the core.

\subsection{Transverse heat flow approximation}

Our method does not allow longitudinal flow of heat. We justify this by noting the large difference in length scale for the transverse and longitudinal temperature variations. The important longitudinal length scale is the modal beat length. This varies with fiber design but it typically falls in the range 3-30 mm in large mode fibers. The transverse length scale is the core diameter which typically lies in the range 20-80 $\mu$m. The ratio of longitudinal to transverse length is of order 100:1. 

\subsection{Narrow line width approximation}

Our model relies on a frequency offset between the light in different transverse modes. The frequency offset is usually in the range 300-3000 Hz. The assumption of periodic behavior implies the set of all allowed frequencies must be separated by multiples of the inverse modulation period. The light in any mode must consist of only these frequencies. The width of each frequency component is assumed to be much smaller than the frequency offset. This narrow line width is obviously not entirely realistic. Nevertheless, as we will discuss below in Sec.~{\ref{sec:narrow}, this method can still be a highly accurate way of treating the problem of mode coupling in relatively narrow band amplifiers.

\subsection{Split-step integration in t and z}

Because it is only necessary to treat one oscillation period under the steady-periodic assumption, we perform the integration in $z$ on a set of time samples of the field that are evenly spaced over one period. At the fiber input we specify $E_s(x,y,0,t)$ and the pump power. Using this information we compute the temperature profiles for each sampling time over the full period. We describe two methods for the integration of the heat equation below. The temperature profile for each time sample is then used to propagate the corresponding time sample of the signal field by $\Delta z$. This is repeated until the end of fiber is reached. This method is used because it makes efficient use of limited computer resources, and because it can be made to run fast. 

We use a split-step, fast-Fourier-transform, beam propagation method (FFT BPM) to perform the $z$ integrations. Split-step methods have been employed in numerical simulations of CW optical diffraction, including guiding in optical fibers, at least since the 1970s\cite{Feit1980_1,Feit1980_2}. One period of the field is discretized into $N_t$ time steps. We use the split-step method by applying it individually to each of the field time slices. Related FFT methods for dispersive propagation are also widely used to model propagation of short light pulses propagating in a single fiber mode\cite{Agrawal}. Although we are also propagating pulses consisting of one beat time, it is best to think of the time slices as widely separated samples that are unaffected by dispersion. We will say more about this in Sec.~\ref{sec:narrow}.

In the split-step method, Eq.~(\ref{eqn:paraxial_wave_eqn_with_gain}) is used to advance each field time sample by $\Delta z$ in three steps. In the first step, linear diffractive propagation is applied to advance the field by $\Delta z/2$ in a homogeneous medium with a refractive index equal to the cladding index. In this step, the beam propagation equation is used, keeping only the diffractive term on the right-hand-side
\begin{equation}
\label{eqn:paraxial_wave_eqn2} \frac{\partial E_s(x,y,z,t)}{\partial z} = \frac{i}{2 k_c}\nabla^2_\perp E_s(x,y,z,t).
\end{equation}
In the second step, the field is advanced by $\Delta z$ without diffraction, keeping only the phases induced by the guiding and thermally induced index, along with the laser gain and linear loss contained in $g(x,y,z,t)$. The propagation equation for this step becomes
\begin{equation}
\label{eqn:paraxial_wave_eqn3} \frac{\partial E_s(x,y,z,t)}{\partial z} =  \frac{i\big[k^2(x,y,z,t)-k_c^2\big]}{2k_c}E_s(x,y,z,t) + g(x,y,z,t) E_s(x,y,z,t).
\end{equation}
The gain term will be described in more detail in the next section. In the third step, linear diffractive propagation is again applied to advance the field by $\Delta z/2$. This method produces errors of order $\mathcal{O}(\Delta z^3)$.

\section{Algorithm}

\begin{figure}[hpb]
\centering
\includegraphics[height=0.95\textheight]{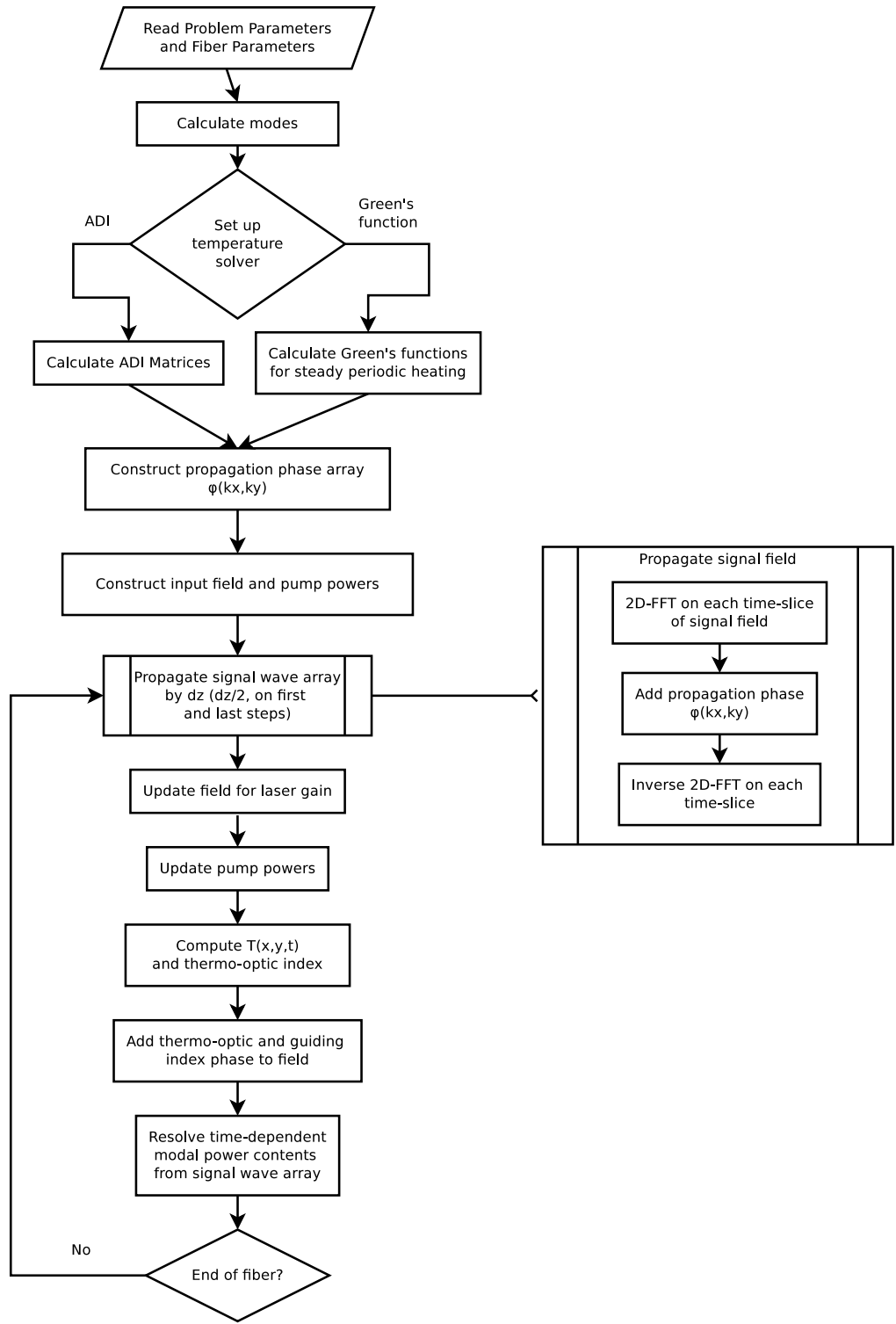}
\caption{\label{fig:algorithm}Flow diagram for split-step FFT propagation using alternating direction implicit integration (ADI) or steady-periodic Greens function method for temperature calculation.}
\end{figure}

These methods are incorporated in our algorithm which is diagrammed in Fig.~\ref{fig:algorithm}. In the following sections, we describe each step of the algorithm in more detail.

\section{Read problem parameters and fiber parameters}

\begin{table}[htp]
\caption{\label{tab:input_params}Model input parameters}
\centering
\begin{tabular}[3]{cccc}\hline
 $L_x$ & $L_y$ & $\Upsilon$&L\\
$N_x$ & $N_y$ & $N_t$ & $\Delta z$\\
 $d_{\rm core}$ & $d_{\rm clad}$ & $d_{\rm dopant}$&$N_{\rm Yb}(x,y)$\\
$\sigma_{s}^{a}$ & $\sigma_{s}^{e}$ & $\sigma_{p}^{a}$ & $\sigma_{p}^{e}$\\
$\lambda_{p}$ & $\lambda_{s}$ &$\tau$ & $\Delta_{\rm sample}$\\
$P_{m,n}^s$ & $P^{p}_{f}$ & $P^{p}_{b}$ & $n_{\rm clad}$ \\
$M_{m,n}^s(t)$ & $M^p(t)$ & $\alpha(x,y)$  & $n_{\rm core}(x,y)$ \\
$\rho$ & ${\rm d}n/{\rm d}T$ & $C$ & $K$\\
$R_{\rm bend}$ &$\gamma_{m,n}$\\
\end{tabular}
\end{table}

The first step in the algorithm is to read in the problem parameters. Table~\ref{tab:input_params} lists them individually. Some of them are standard silica parameters. For example: thermal conductivity $K=1.38$ W/m-K; specific heat $C=703$ J/kg-K; density $\rho=2201$ kg/m$^3$; thermo-optic coefficient ${\rm d}n/{\rm d}T=1.2\times 10^{-5}$ K$^{-1}$. Usually the pump wavelength is $\lambda_p=976$ nm, and the upper state ion lifetime is $\tau=901$ $\mu$s.

We define our problem at each $z$-location on an $(x,y,t)$ grid. The grid is equally spaced in $(x,y,t)$, with number of points typically $(N_x=64, N_y=64, N_t=64)$. The spatial grid spans a domain of size $ L_x \times L_y $ with the core at its center, where $L_x$ and $L_y$ are typically $\approx 3\times d_{\rm core}$. We choose as the thermal boundary condition a fixed temperature on the domain boundary. We assume there is a fixed frequency offset between modes equal to $\Delta\omega$. The period $\Upsilon$ is defined as one cycle of the beat frequency, $\Upsilon=2\pi/\Delta \omega$.  The grid step sizes are then
\begin{eqnarray}
\Delta x &=& L_x / (N_x - 1)\\
\Delta y &=& L_y / (N_y - 1)\\
\Delta t &=& \Upsilon / N_t.
\end{eqnarray}

The parameters defining the doping profile $N_{\rm Yb}(x,y)$, the refractive index profile $n_{\rm core}(x,y)$ and the linear loss $\alpha(x,y)$ vary with $(x,y)$ position. We usually use super-Gaussian profiles for these, with super-Gaussian coefficient of order 40. The FWHM values for the super-Gaussian diameters are $d_{\rm core}$ and $d_{\rm dopant}$. The pump is confined to the pump cladding of diameter $d_{\rm clad}$. The absorption and emission cross sections for pump and signal are $\sigma_p^a$, $\sigma_p^e$, $\sigma_s^a$, $\sigma_s^e$.

The forward and backward time-averaged input pump powers, $P^{p}_{f}$ and $P^{p}_{b}$, may be periodically modulated by specifying a pump modulation function $M^p(t)$. Similarly, each signal mode LP$_{m,n}$ with time averaged power $P_{m,n}^s$ can be modulated by specifying its modulation function $M_{m,n}^s(t)$. The $\gamma_{m,n}$ are mode specific losses (non heating).

The fiber length is $L$, and the fiber is bent to a radius of $R_{\rm bend}$. The integration step size $\Delta z$ is typically set to a few microns. We store information about the field every $\Delta_{\rm sample}$ integration steps. Typically the stored information includes the local time sequences for the mode content, total signal power, effective area of the signal, and pump power.

\section{Calculate modes}

The next step of the algorithm is to compute the signal mode profiles. Here, we present a general method capable of handling bent fiber and arbitrary refractive index profiles of low contrast. This procedure is from Marcuse\cite{Marcuse}. It is not a vector method so it is not appropriate for air holes or other guiding structure with high index contrast. Vector adaptations that permit high index contrast are available, but we do not discuss them here. 

On the rectangular grid $(0<x<L_x)$ by $(0<y<L_y)$, the functions
\begin{equation}\label{eqn:basis_set}
\phi_{\mu\nu}=\sqrt{\frac{4}{L_xL_y}}\sin\biggl(\frac{\pi}{L_x}\mu x\biggr)\sin\biggl(\frac{\pi}{L_y}\nu y\biggr)
\end{equation}
with $\mu$ and $\nu$ being integers, form a complete, orthonormal set, normalized so their areal integral is unity. Any transverse mode $\psi (x,y,z)$ can be written
\begin{equation}\label{eqn:modesinexp}
\psi (x,y,z)=e^{-i\beta z}\sum_{\mu=1}^{\infty}\sum_{\nu=1}^{\infty}C_{\mu\nu}\phi_{\mu\nu}(x,y)
\end{equation}
where $\beta$ is the propagation constant for that mode, $C_{\mu \nu}$ is the mode's eigenvector (coefficients of the basis set $\phi_{\mu \nu}$ from Eq.~(\ref{eqn:basis_set})) and $\psi$ is a solution of the Helmholtz equation
\begin{equation}\label{eqn:helmholtz}
\frac{\partial^2\psi}{\partial x^2}+\frac{\partial^2\psi}{\partial y^2}+\frac{\partial^2\psi}{\partial z^2}+n^2(x,y)k_{\circ}^2\psi=0
\end{equation}
where $k_{\circ}=\omega_c/c$. We set an upper limit on the number of sin-sin terms to include in the expansion of $\psi$ ($M_x$ and $M_y$), and substituting this expansion into the Helmholtz equation yields
\begin{equation}
\sum_{\mu=1}^{M_x}\sum_{\nu=1}^{M_y}C_{\mu\nu}\biggl[n^2(x,y)k_{\circ}^2-\beta^2-\pi^2\biggl(\frac{\mu^2}{L_x^2}+\frac{\nu^2}{L_y^2}\biggr)\biggr]\phi_{\mu\nu}(x,y)=0.
\end{equation}
Next multiply this equation by $\phi_{\mu^{\prime}\nu^{\prime}}$ and integrate it over $x$ and $y$. We define $n_{\rm eff}=\beta/k_{\circ}$ and we divide the equation by $k_{\circ}^2$ and add and subtract $n_{\rm clad}^2$ under the summation sign to obtain
\begin{equation}\label{eqn:eigvaleq}
\sum_{\mu=1}^{M_x}\sum_{\nu=1}^{M_y}A_{\mu^{\prime}\nu^{\prime} ,\; \mu\nu}\:C_{\mu\nu}=(n_{\rm eff}^2-n_{\rm clad}^2)C_{\mu^{\prime}\nu^{\prime}}.
\end{equation}
The $A$ matrix is
\begin{eqnarray}\nonumber\label{eqn:AA}
A_{\mu^{\prime}\nu^{\prime} ,\; \mu\nu}&=&\int_0^{L_x}\int_o^{L_y} {\rm d}x {\rm d}y \Bigl\{\bigl[n^2(x,y)-n_{\rm clad}^2\bigr]\phi_{\mu^{\prime}\nu^{\prime}}(x,y)\phi_{\mu \nu}(x,y)\Bigr\}\\
&&-\frac{\pi^2}{k_{\circ}^2}\biggl(\frac{\mu^2}{L_x^2}+\frac{\nu^2}{L_y^2}\biggr)\delta_{\mu ,\mu^{\prime}}\delta_{\nu ,\nu^{\prime}}.
\end{eqnarray}
Equation (\ref{eqn:eigvaleq}) is an eigenvalue equation of the form
\begin{equation}
AC=\bar{n}^2 C
\end{equation}
to be solved for eigenvalues $\bar n$ and eigenvectors $C$. The solution eigenvectors are substituted in Eq. (\ref{eqn:modesinexp}) to compute the eigenmodes. Then the effective refractive index  $n_{\rm eff}$ of mode LP$_{m,n}$ is related to its eigenvalue $\bar{n}_{m,n}$ by
\begin{equation}\label{eqn:n_eff}
n_{\rm eff}(m,n)=\sqrt{\bar{n}_{m,n}^{\;2} + n_{\rm clad}^2}.
\end{equation}
The beat length between LP$_{01}$ and LP$_{11}$ can be found using their values for $n_{\rm eff}$.

For simple guiding index shapes such as a step index in unbent fiber, the integrals in Eq. (\ref{eqn:AA}) can be integrated analytically, otherwise they can be evaluated numerically. We use numerical integration because of its generality. 

The modes computed this way are the low power modes that do not take into account a nonuniform temperature or any other irradiance-dependent index changes. These low power basis modes will be used to construct input fields and to analyze propagating fields into modes.

\subsection{Refractive index profile}

The method described is general enough to consider arbitrarily-shaped refractive index profiles. For simplicity, we typically use a top hat-like two-dimensional super-Gaussian of high order to construct our refractive index and doping profiles. The super-Gaussian index profile is computed using
\begin{equation}\label{eqn:super-gaussian}
n(x,y) = n_{\rm clad} + (n_{\rm core}-n_{\rm clad})\times \exp\bigg(-\ln 2 \Big\{ \big[2(x-x_0)/d_{\rm core}\big]^2 + \big[2(y-y_0)/d_{\rm core}\big]^2 \Big\}^{S} \bigg),
\end{equation}
where $n_{\rm clad}$ and $n_{\rm core}$ are the refractive indices of the cladding and core, $(x_0,y_0)$ are the coordinates of the core center, $d_{\rm core}$ is the core diameter, and $S$ is the super-Gaussian coefficient (we typically use $S=40$).

We approximate the effects of bending the fiber by adding a refractive index ramp $n_{\rm bend}$ to $n(x,y)$ in the plane of the bend so the index is higher on the outside of the bend. If bending is in the $\hat{x}$ plane, this added ramp can be written
\begin{equation}
n_{\rm bend}(x,y) = n(x,y)(x-x_0)/R_{\rm bend}.
\end{equation}
Bend losses are calculated using the method of Marcuse\cite{Marcuse1976}. In the numerical model bend and other losses from the core are enforced by included an absorbing layer near the grid boundary.

\subsection{Normalization}

In order to easily relate our calculated modes to power in watts, we introduce a normalization factor $N_{m,n}$ for each mode constructed from Eq.~(\ref{eqn:modesinexp}), where $N_{m,n}$ is the value which satisfies
\begin{equation}
\frac{c\epsilon_0}{2} \int {\rm d}x {\rm d}y\ n(x,y) \Big[N_{m,n}\psi_{m,n}(x,y)\Big]^2 = 1 {W}
\end{equation}
The normalized mode $u_{m,n}(x,y)$ is defined by
\begin{equation}\label{eqn:normalized_modes}
u_{m,n}(x,y) = N_{m,n} \psi_{m,n}(x,y).
\end{equation}

\section{Set up temperature solver}\label{sec:heat_stuff}

Next the temperature profile must be solved for each time point in the period $\Upsilon$ using the periodic heat source $Q$. In an isotropic homogeneous medium, the heat equation is 
\begin{equation}\label{eqn:heat_equation}
\rho C \frac{\partial T}{\partial t} = Q + K \biggl( \frac{\partial^2 T}{\partial x^2} + \frac{\partial^2 T}{\partial y^2} \biggr)
\end{equation}
where $T=T(x,y,t)$ is the temperature, $Q=Q(x,y,t)$ is the heating source, $K$ is the scalar thermal conductivity, $\rho$ is the density and $C$ is the specific heat capacity. No $\partial^2 / \partial z^2$ term is included in Eq.~(\ref{eqn:heat_equation}) because we don't allow longitudinal heat flow. We consider only the thermal boundary condition of a fixed temperature on the $(x,y)$ grid boundary. Alternative boundary conditions can be dealt with by modifying our procedure, but we do not discuss them here.

We will describe two methods of solving Eq.~(\ref{eqn:heat_equation}) for $T$. The first method, the Green's function method for steady-periodic heating, involves computing the temperature over the $(x,y)$ grid as a sum of independent contributions, one from each heated grid point. The second method, the alternating direction implicit (ADI) integration, involves stepping in time via successive matrix multiplications involving all pixels contributing together.

\subsection{Calculate Green's functions for steady-periodic heating}\label{subsec:greens_fcn}

To use the Green's function method we must first compute the steady-periodic Green's function which gives the temperature contributions over the entire $(x,y)$ grid due to the periodically heated point $(x^{\prime},y^{\prime})$. These functions describe the temperature rise due to a steady-periodic heat source at a single modulation frequency. Formulations of the Green's functions for different boundary conditions are detailed in \cite{Cole}. For temperatures clamped at all four sides of the grid, the complex valued Green's function is
\begin{equation}
G(x,y,\omega|x^\prime,y^\prime) = \sum_{n=0}^{\infty} F_n(y,y^\prime)P_n(x,x^\prime,\omega)
\end{equation}
\begin{equation}
F_n(y,y^\prime) = \frac{1}{WK}\sin\biggl( \frac{n\pi}{W} y \biggr) \sin \biggl( \frac{n\pi}{W} y^\prime \biggr)
\end{equation}
\begin{eqnarray*}
P_n(x,x^\prime,\omega) & = & \bigg\{ \exp\big[-\sigma_n(2H-|x-x^\prime|)\big] - \exp\big[-\sigma_n(2H-x-x^\prime)\big] \\
& & + \exp\big[-\sigma_n|x-x^\prime|\big] - \exp\big[-\sigma_n(x+x^\prime)\big] \bigg\} \bigg/ \\
& & \sigma_n \Big(1 - \exp\big[-2\sigma_n H\big]\Big)
\end{eqnarray*}
where
\begin{equation}
\sigma_n^2 = \biggl( \frac{n\pi}{W} \biggr)^2 + i \omega \frac{\rho C}{K},
\end{equation}
$\omega$ is the frequency of the heat, ($0\le x\le H$), and ($0\le y \le W$).

We compute these functions for several harmonics, making sure the time grid is sufficient to resolve them. That is, we find $G(x,y,\omega_m|x^\prime,y^\prime)$ for $\omega_m=(0,1,2,3,...) \times \Delta\omega$ where $\Delta\omega$ is the specified frequency offset. The use of this set of Green's functions to compute $T(x,y,t)$ is described in Sec.~\ref{subsec:greens_temperature}.

\subsection{Calculate ADI matrices}\label{subsec:ADI_matrices}

An alternative, more general method of solving the heat equation is the alternating direction implicit (ADI) method\cite{NR}. This method does not require periodicity. When it is applied to a steady-periodic problem, it requires many iterations to match the steady-periodic requirement, with each iteration operating on the previous iteration's data. This sequential process means a simple parallel computing scheme is difficult to implement for ADI integration.

The derivation of the equations for this method begins by using finite difference definitions of the partial derivatives in Eq.~(\ref{eqn:heat_equation}). The numerical integration of the heat equation by time step $\Delta t$ is split into two halves. We first integrate explicitly in $y$ and implicitly in $x$ for one half time step $\Delta t/2$, then integrate explicitly in $x$ and implicitly in $y$ for another half time step $\Delta t/2$.

We start with the combined expression of implicit-$x$/explicit-$y$ integration
\begin{eqnarray}
\nonumber T_{x,y}^{t+\Delta t/2} - T_{x,y}^t & = & \big( T_{x+\Delta x,y}^{t + \Delta t/2} - 2T_{x,y}^{t + \Delta t/2} + T_{x-\Delta x,y}^{t+\Delta t / 2} \big) \frac{\lambda_x}{2} \\ 
& & + \big( T_{x,y+\Delta y}^{t} - 2T_{x,y}^{t} + T_{x,y-\Delta y}^{t} \big) \frac{\lambda_y}{2} \\
& & \nonumber +\frac{\Delta t}{2 \rho C} Q^{t+\Delta t /4}
\end{eqnarray}
or, rewritten in matrix form,
\begin{eqnarray}
T^{t+\Delta t/2} A_x = B_y T^{t} + \frac{\Delta t}{2\rho C} Q^{t+\Delta t/4}\\
T^{t+\Delta t/2} = B_y T^{t}A_x^{-1} + \frac{\Delta t}{2\rho C} Q^{t+\Delta t/4}A_x^{-1},
\end{eqnarray}
where $A_x$ is the matrix for implicit integration in the x-direction, and $B_y$ is the matrix for explicit integration in the y-direction. These matrices are tridiagonal, and $A_x$ has size $(N_x-2)\times(N_x-2)$ while $B_y$ has size $(N_y-2)\times(N_y-2)$. The matrices are
\begin{equation}\label{eqn:implicit_x}
A_x=\left[\begin{array}{ccccccc}
(1+\lambda_x)&-\lambda_x/2&0&\cdots&0&0&0\\
-\lambda_x/2&(1+\lambda_x)&-\lambda_x/2&\cdots&0&0&0\\
\vdots&\vdots&\vdots&\ddots&\vdots&\vdots&\vdots\\
0&0&0&\cdots&-\lambda_x/2&(1+\lambda_x)&-\lambda_x/2\\
0&0&0&\cdots&0&-\lambda_x/2&(1+\lambda_x)\\
\end{array}\right],
\end{equation}

\begin{equation}\label{eqn:explicit_y}
B_y=\left[\begin{array}{ccccccc}
(1-\lambda_y)&\lambda_y/2&0&\cdots&0&0&0\\
\lambda_y/2&(1-\lambda_y)&\lambda_y/2&\cdots&0&0&0\\
\vdots&\vdots&\vdots&\ddots&\vdots&\vdots&\vdots\\
0&0&0&\cdots&\lambda_y/2&(1-\lambda_y)&\lambda_y/2\\
0&0&0&\cdots&0&\lambda_y/2&(1-\lambda_y)\\
\end{array}\right].
\end{equation}
Matrix $A_x$ has diagonal elements $(1+\lambda_x)$ and off-diagonal elements $-\lambda_x/2$, while matrix $B_y$ has diagonal elements $(1-\lambda_y)$ and off-diagonal elements $\lambda_y/2$, where

\begin{eqnarray}
\label{eqn:lambda_x}\lambda_x &=& \frac{K\Delta t}{\rho C \Delta_x^2} \\
\label{eqn:lambda_y}\lambda_y &=& \frac{K\Delta t}{\rho C \Delta_y^2}.
\end{eqnarray}

In the second half time step, implicit/explicit integration order is reversed
\begin{eqnarray}
\nonumber T_{x,y}^{t+\Delta t} - T_{x,y}^{t + \Delta t/2} &=& \big(T_{x+\Delta x,y}^{t+\Delta t/2} - 2T_{x,y}^{t+\Delta t/2} + T_{x-\Delta x,y}^{t + \Delta t/2} \big) \frac{\lambda_x}{2}\\
& &  + \big( T_{x,y+\Delta y}^{t+\Delta t} - 2T_{x,y}^{t+\Delta t} + T_{x,y-\Delta y}^{t + \Delta t} \big) \frac{\lambda_y}{2} \\
& & \nonumber + \frac{\Delta t}{2\rho C} Q^{t+3\Delta t/4}.
\end{eqnarray}
Or, rewritten in matrix form,
\begin{eqnarray}
\nonumber A_y T^{t+\Delta t} &=& T^{t+\Delta t/2}B_x + \frac{\Delta t}{2\rho C}Q^{t+3\Delta t/4}\\
T^{t+\Delta t} &=& A_y^{-1}T^{t+\Delta t/2}B_x + \frac{\Delta t}{2\rho C}A_y^{-1}Q^{t+3\Delta t/4},
\end{eqnarray}
where $A_y$ and $B_x$ follow a definition similar to Eqs.~\ref{eqn:implicit_x} and \ref{eqn:explicit_y}, with $\lambda_x$ swapped with $\lambda_y$ and redimensioned as appropriate.

Combining the two half time step integrations, using the ADI to integrate by one full time step $\Delta t$ becomes
\begin{equation}\label{eqn:adi_step}
T^{t+\Delta t} = A_y^{-1}B_yT^tA_x^{-1}B_x + \frac{\Delta t}{2 \rho C}A_y^{-1} \bigg( Q^{t+\Delta t/4}A_x^{-1}B_x + Q^{t+3\Delta t/4} \bigg),
\end{equation}
where, because we only have $Q$ defined on integral numbers of time steps, we use linear interpolation to find $Q^{t+\Delta t/4}$ and $Q^{t + 3\Delta t/4}$ from the heat at $t$ and $t+\Delta t$. This method is converged to $\mathcal{O}(\Delta t^2)$.

\section{Construct propagation phase array}

Linear diffractive propagation that comprises the first and third steps in the split-step propagation is best evaluated in $(k_x,k_y,z,t)$ space. Equation (\ref{eqn:paraxial_wave_eqn2}) can be transformed to this space by inserting the following form of $E_s(x,y,z,t)$
\begin{equation}
E_s(x,y,z,t)=\frac{1}{{2\pi}}\int_{-\infty}^{\infty}\int_{-\infty}^{\infty}E_s(k_x,k_y,z,t)e^{ik_x x}e^{ik_y y}{\rm d}k_x{\rm d}k_y.
\end{equation}
The transformed equation is
\begin{equation}
\frac{\partial E_s(k_x,k_y,z,t)}{\partial z}=-i\frac{k_x^2}{2k_c}E_s(k_x,k_y,z,t)-i\frac{k_y^2}{2k_c}E_s(k_x,k_y,z,t).
\end{equation}
Advancing the field $E(k_x,k_y,z,t)$ by $\Delta z/2$ consists of shifting the phase of each ($k_x$,$k_y$) plane wave component by 
\begin{equation}\label{eqn:phase_advance}
\phi(k_x,k_y) =- \frac{\Delta z}{2} \frac{(k_x^2 + k_y^2)}{2k_c}.
\end{equation} 
The inverse two-dimensional Fourier transform is used to convert $E_s(k_x,k_y,z,t)$ back to $E_s(x,y,z,t)$. In practice, fast Fourier transforms (FFTs) are used to efficiently convert fields between $(x,y,z,t)$ and $(k_x,k_y,z,t)$ spaces. 

\section{Construct input signal field and pump powers}

\subsection{Signal}

We use the set of normalized modes $u_{m,n}(x,y)$ defined in Eq.~(\ref{eqn:normalized_modes}) to construct the input signal field 
\begin{equation}
E_s (x,y,z,t)\Big|_{z=0} = \sum_{\rm modes} \sqrt{P^{\;s}_{m,n}}\ u_{m,n}(x,y)M_{m,n}^{\;s}(t)
\end{equation}
where $P^{\;s}_{m,n}$ is the power in the $({m,n})^{\rm th}$ mode and $M_{m,n}^{\;s}(t)$ the periodic modulation function for that mode. For example, to model an amplifier with LP$_{01}$ populated by unshifted light, and LP$_{11}$ populated by light detuned by $\Delta\omega$, we set $M_{0,1}^s(t)=1$, $M_{1,1}^s(t)=\exp(-i\Delta\omega t)$. If instead we wish to amplitude modulate the signal in both modes with a simple sinusoidal envelope, we make each modulation function $M_{m,n}^s(t)=1+0.25a\cos{(\Delta\omega t)}$ where $a$ is the small peak-to-peak power modulation. We can also use more complicated modulations, as long as they are periodic. For more complicated modulation we must include an adequate number of harmonics in the Green's function temperature solver.

\subsection{Pump}

Defining a co-propagating pump input is simple, but since we start the integration at the signal input end, specifying the counter-propagating pump that gives the target pump at the opposite end can be more difficult, especially if amplitude modulations are involved. To keep the two pumps separate we define two quantities $P^{p}_{f}(z,t)|_{z=0}$ and $P^{p}_{b}(z,t)|_{z=0}$ for forward- and backward-propagating pumps.

To generate modulated input pump powers, we follow a technique similar to the signal modulation considered above for each pump. The model is capable of treating an arbitrary periodic pump modulation, provided we include an adequate number of harmonics if solving the heat equation using the Green's function method.

\section{Propagate signal field}

To reiterate the propagation described earlier, the signal is propagated by Fourier transforming each time slice of the field $E_s(x,y,z,t)$ to $E_s(k_x,k_y,z,t)$ using 2D-FFTs. The phase of each plane wave component is advanced by the propagation phase $\phi(k_x,k_y)$ given in Eq.~(\ref{eqn:phase_advance}), and the inverse 2D-FFTs of $E_s(k_x,k_y,z,t)$ gives the updated field $E_s(x,y,z,t)$. If the propagation step is not the first or last half-step in the fiber, the two consecutive half-step propagations are combined by advancing the phase by $2\phi(k_x,k_y)$.

\section{Update the field for laser gain}

In the non diffractive part of the split-step procedure we update the signal field by adding the guiding and thermally induced phases plus any laser gain and linear loss. To compute the gain we find the upper state population profile in $(x,y,t)$ from the signal irradiance profile, pump power, and doping profile, and use it to compute the signal gain and pump loss. The (steady-state) upper state population is
\begin{equation}\label{eqn:upper_state}
n_u(x,y,t) = \frac{P_p \sigma_p^a / h \nu_p A_p + I_s \sigma_s^a / h \nu_s}{P_p(\sigma_p^a + \sigma_p^e)/h \nu_p A_p + I_s(\sigma_s^a + \sigma_s^e)/h\nu_s + 1/\tau}
\end{equation}
where $I_s=I_s(x,y,t)$ is the signal irradiance, $\nu_s$ and $\nu_p$ are the signal and pump frequencies, $\sigma_s^a$ and $\sigma_s^e$ are the signal absorption and emission cross sections, $\sigma_p^a$ and $\sigma_p^e$ are the pump absorption and emission cross sections, $\tau$ is the ion upper-state lifetime, $P_p$ is the pump power, and $A_p$ is the area of the pump cladding. We assume that the cross sections and lifetime are independent of temperature. The effective ion lifetime at typical amplifier powers is a few micro seconds so at modulation frequencies of order 1 kHz the steady state solution is appropriate.

The change in signal field is given by
\begin{equation}\label{eqn:signal_gain}
\frac{\partial E_s(x,y,t)}{\partial z} =  \frac{1}{2} \Big[-\sigma_s^a + (\sigma_s^a+\sigma_s^e)n_u(x,y,t)\Big]N_{\rm Yb}(x,y)E_s(x,y,t)-\frac{1}{2}\alpha(x,y) E_s(x,y,t),
\end{equation}
where $N_{\rm Yb}(x,y)$ is the Yb$^{3+}$ doping profile, $\alpha(x,y)$ is the linear absorption coefficient. This method is general enough to treat an arbitrarily-shaped doping profile, but we typically consider super-Gaussian doping profiles similar to the one defined in Eq. (\ref{eqn:super-gaussian}). The linear absorption coefficient $\alpha$ can be non-uniform in $(x,y)$ to accommodate a photodarkening model. We assume that all the power absorbed due to $\alpha(x,y)$ is turned into heat. Referring to Eq.~(\ref{eqn:paraxial_wave_eqn3}), the laser gain and loss term $g(x,y,t)E_s(x,y,t)$ is given by the right hand side of Eq.~(\ref{eqn:signal_gain}).

\section{Update pump powers}

We include both forward- and backward-going pumps, with the total pump power given by
\begin{equation}
P_p = P^{p}_{f} + P^{p}_{b}.
\end{equation}
$P_p$ is assumed to be uniformly distributed across the pump cladding. The change in the pump power is computed directly from the ion inversion, rather than from the signal increment. This allows us to include linear signal loss and fluorescence loss correctly. The total change in pump power is given by
\begin{equation}
\frac{{\rm d}P_p}{{\rm d}z} = \frac{P_p}{A_p} \int \int \Big[(\sigma_p^a+\sigma_p^e)n_u(x,y)-\sigma_p^a\Big] N_{\rm Yb}(x,y) {\rm d}x {\rm d}y.
\end{equation}
The loss is apportioned between the forward and backward pumps according to
\begin{equation}
\frac{{\rm d}P^{p}_{f}}{{\rm d}z} = \frac{{\rm d}P_p}{{\rm d}z} \frac{P^{p}_{f}}{P_p}
\end{equation}
\begin{equation}
\frac{{\rm d}P^{p}_{b}}{{\rm d}z} = -\frac{{\rm d}P_p}{{\rm d}z} \frac{P^{p}_{b}}{P_p}.
\end{equation}

\section{Compute T(x,y,t) and thermo-optic index}\label{sec:temperature_calc}

For the computation of $T(x,y,t)$ we use either the Green's function method or the ADI method. We calculate the heat deposition rate from the absorbed pump and the quantum defect according to
\begin{equation}
Q(x,y,t) = N_{\rm Yb}(x,y)\Bigl[ \frac{\nu_p-\nu_s}{\nu_p} \Bigr] \Bigl[ \sigma_p^a - (\sigma_p^a+\sigma_p^e)n_u(x,y,t) \Bigr] \frac{P_p(t)}{A_p} + \alpha(x,y) I_s(x,y,t),
\end{equation}
where the upper state population $n_u(x,y,t)$ is given by Eq.~(\ref{eqn:upper_state}).

\subsection{Green's function method}\label{subsec:greens_temperature}

The heat as a function of time over the full cycle at each $(x^{\prime},y^{\prime})$ pixel is resolved into its Fourier components $\omega_m=(0,1,2,3...) \times \Delta\omega$ by performing a temporal Fourier transform on $Q(x^{\prime},y^{\prime},t)$
\begin{equation}
q(x^\prime,y^\prime,\omega_m) = \Delta x \Delta y \sum_{i=0}^{N_t-1}  Q(x^\prime,y^\prime,t_i)\ \exp{(-i\omega_m t_i)}.
\end{equation}
Here, $q(x^\prime, y^\prime, \omega_m)$ is a complex coefficient that includes the phase as well as the amplitude of the heat deposition. These $q(x^\prime,y^\prime,\omega_m)$ values are used to weight the Green's functions in computing the steady-periodic temperature over the entire transverse grid. We always include frequencies $(0,1)\times \Delta\omega$, and we add higher frequency terms as needed for convergence. The time grid must be fine enough to resolve the highest frequency. Higher frequency terms are usually necessary only above the mode instability threshold. 

Using the Green's functions computed as described in Sec.~\ref{subsec:greens_fcn}, the temperature $T(x,y,t)$ is found by summing the contributions of each $q(x^\prime,y^\prime,\omega_m)$ according to
\begin{equation}
T(x,y,t) = \textrm{Real} \Bigg[ \sum_{m=0}^{max}\;\;\sum_{x^\prime,y^\prime} q(x^\prime, y^\prime, \omega_m)\ G(x,y,\omega_m|x^\prime,y^\prime)\ \exp(i\omega_m t) \Bigg].
\end{equation}

\subsection{ADI Method}

In Sec.~\ref{subsec:ADI_matrices}, we described a method of integrating the heat equation using the ADI method. This method uses Eq.~(\ref{eqn:adi_step}) to integrate one time step of $\Delta t$. For steady-periodic heating, we enforce the steady-state criterion by integrating for several periods, reusing the heat $Q(x,y,t)$ from one period to the next. We terminate this process when the difference in temperatures between periods has reached an acceptably small residual.

The ADI method is more general than the steady-periodic Green's function method, because it does not require periodicity. It can be used to model transient heating, for example, if the steady-periodic condition is not enforced. It is unconditionally stable, but its accuracy suffers if the time-step is too large.

\section{Add thermo-optic and guiding index phases to the field}

In the non diffraction portion of the split-step propagation we also advance the phase of the field to account for the guiding index plus the thermal index over $\Delta z$ using
\begin{equation}
\phi(x,y,t) =  \Delta z \frac{k^2(x,y,t)-k_c^2 }{2k_c}.
\end{equation}
The phase can be rewritten using Eq.~(\ref{eqn:core_profile}) as
\begin{equation}
\phi(x,y,t) = \Delta z \frac{\omega_c}{c}\Bigg( [n(x,y,t) - n_{\rm clad}] + \frac{[n(x,y,t) - n_{\rm clad}]^2}{2 n_{\rm clad}} \Bigg).
\end{equation}
We write it in this form merely to show that the phase is $(\omega_c\Delta z \Delta n/c)$ plus a small correction from the quadratic term.

\section{Resolve time-dependent modal power contents}

We use the normalized modes defined in Eq.~(\ref{eqn:normalized_modes}) to resolve the content of the signal field $E_s(x,y,z,t)$ into fiber modes. We compute the inner product of the field and the normalized mode from
\begin{equation}\label{eqn:mode_amp}
F_{m,n}(z,t) = \frac{\int\!\!\!\!\int {\rm d}x {\rm d}y\  E_s(x,y,z,t) u_{m,n}(x,y)}{\int\!\!\!\!\int {\rm d}x {\rm d}y\  \big[u_{m,n}(x,y)\big]^2},
\end{equation}
where $F_{m,n}(z,t)$ is a complex quantity. The mode amplitude $F_{m,n}(z,t)$ is converted to power in watts using
\begin{equation}\label{eqn:mode_power}
P_{m,n}(z,t) = \Big | F_{m,n}(z,t)\Big |^2.
\end{equation}

\subsection{Spectral analysis}

To find the frequency spectrum of mode $(m,n)$ we Fourier transform $F_{m,n}(z,t)$ from time to frequency. The allowed frequencies for the periodic function $F_{m,n}(z,t)$ are the $\omega_m$.

\subsection{Compute effective area}

We also compute the effective area of the signal field using
\begin{equation}\label{eqn:a_eff}
A_{\rm eff}(z,t) = \frac{\Big[\int\!\!\!\int \big|E_s(x,y,z,t)\big|^2 {\rm d}x {\rm d}y \Big]^2}{\int\!\!\!\int \big|E_s(x,y,z,t)\big|^4 {\rm d}x {\rm d}y}.
\end{equation}

\section{An example computation}

In Fig. \ref{fig:surface} we show a sample result produced by the model. The surface shows the power in LP$_{11}$ over one LP$_{01}$-LP$_{11}$ beat length at the input end of an amplifier versus $t$ and $z$. The motion of the surface ridges reflect the change in phase of LP$_{11}$ relative to LP$_{01}$ due to the difference in propagation constants for the two modes. In this example the LP$_{11}$ seed light is Stokes shifted by 600 Hz relative to LP$_{01}$, but the details of the fiber are unimportant for this illustration since all large mode fiber amplifiers produce qualitatively similar surfaces. A small fraction of the gain is due to laser gain but most is due to thermally induced mode coupling.

The offset frequency that produces the highest mode coupling gain varies along the length of the fiber due to the changing shape of the heat profile. We integrate a set of frequencies and powers over the full length of the fiber to find the lowest threshold. When operating near or below threshold the gain curve has a well defined frequency of highest gain even though the shape of the gain varies somewhat along the fiber due to changing heat profiles.
\begin{figure}[htbp]
\includegraphics[width=4.5in]{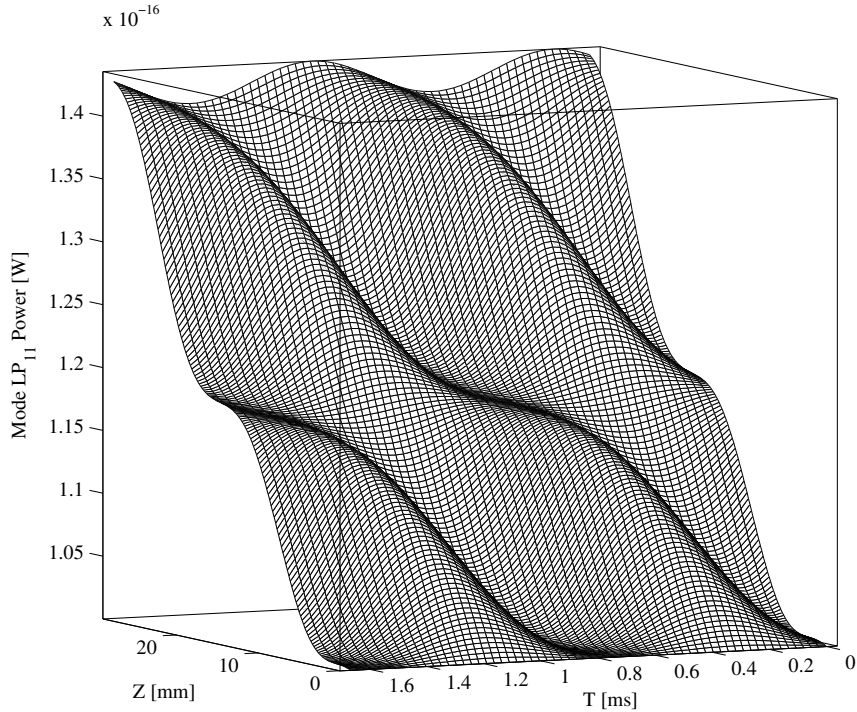}
\caption{\label{fig:surface}Power in LP$_{11}$ versus time (T) and distance (Z) at the input end of the amplifier similar to the one specified in \cite{smithsmith2012a}. This illustrates the growth of the low power mode over one beat cycle and over one beat length. The pump is unmodulated in this example.}
\end{figure}

\section{A narrow band width model for broad band light?}\label{sec:narrow}

Our model assumes that inter modal frequency shifts of order 1 kHz are responsible for thermal mode coupling. How is it possible that a model based on such small shifts can be applicable to light that has much broader line widths? Few seed lasers have sub kHz line widths, and in fact, the seed light is often intentionally phase modulated to widths of several GHz to combat SBS. Our answer is that even broad band light can be frequency shifted as a whole by 1 kHz. Further, for phase modulated light the beat between two waves with identical, but frequency shifted spectra is just as strong as the beat between two frequency shifted monochromatic waves. Additionally, the frequency shift in our model is caused by coupling two modes via a moving thermal grating. The Doppler frequency shift induced by this process shifts the entire spectrum of the scattered light so the interference between modes has full strength. This means that both in the mode coupling process and in the mode beating process responsible for heating, light with a broad spectrum behaves the same as monochromatic light. Also, a phase modulation of several GHz has little impact on the ion population evolution. The steady state expression for the upper state population is still accurate.  

This equivalence of narrow and broad band light breaks down if the two interfering fields shift out of time synchronization. This can happen for large band widths because different modes have slightly different group velocities. The difference in modal propagation times is of order 1 ps/m, so an estimate of the line width limit is 100 GHz for a typical fiber amplifier. However, this is substantially larger than typical SBS suppressing line widths.

If the signal light consists of a train of short pulses, and the pulse separation is shorter than a few micro seconds, the ion population responds primarily to the average power. Slow modulations of the average power, in the kHz range, are still effective in driving the thermal grating that leads to mode coupling. If the pulse width is more than a ps or so the pulses in different modes will stay synchronized well enough to produce strong inter modal interference.

\section{Sources of frequency offset HOM seed light}

It is likely that in most amplifiers an amplitude or spectral modulation of the pump light\cite{smithsmith2012a}, combined with accidental injection of a small fraction of seed light into LP$_{11}$, produces the initial frequency shifted light in LP$_{11}$. Alternatively, amplitude modulation of the seed, plus accidental injection of a fraction into LP$_{11}$ could provide the initial light. If both of these modulations can be sufficiently suppressed, the initial light would be provided by quantum noise. This unavoidable initial light would produce the highest possible instability threshold. Our model does not incorporate a true quantum noise model. Instead we estimate the starting noise power from a classical stochastic electrodynamics noise model\cite{Milonni} often used in estimating thresholds for lasing or for Raman gain. The noise level is set to half a photon per mode. In our case there is a single transverse mode and a gain bandwidth of approximately 1 kHz. The expression for starting power is the same as for Raman noise power, 
\begin{equation}\label{eq.qnoise}
P=h\nu \Delta\nu=(6.6\times 10^{-34})(2.8\times 10^{14})\Delta \nu=1.8\times 10^{-19}\Delta \nu
\end{equation}
which gives $10^{-16}$ W for a bandwidth of 0.5 kHz and a wavelength of 1060 nm. This is the minimum starting power in the higher order mode. The gain process will pick out from the sea of quantum noise the light that best matches the light in LP$_{01}$ both in amplitude and phase over the beat cycle. This is the light with highest gain. There will be fluctuations in frequency and power, but the average power within the gain bandwidth will be well approximated by Eq. (\ref{eq.qnoise}).

\section{Desktop computer implementation}

\subsection{Memory requirements}

Storing the full, four-dimensional double-precision, complex $[64\times 64\times 64\times L/\Delta z]$ array of the signal field would require on the order of $1-10$ terabytes in a meter long amplifier with step sizes of a few microns. Therefore, we do not store $E$ fields at each step. We only store computed properties such as modal content $F_{m,n}(z,t)$, total signal power $I(z,t)$, and effective area $A_{\rm eff}(z,t)$ at positions separated by $(\Delta_{\rm sample} \times \Delta z)$. This reduces the amount of memory required to at most a few gigabytes.

\subsection{Parallel computing}

A substantial portion of the model's run time is spent solving the thermal problem. This makes the Green's function method attractive because it is generally much faster than the ADI method. One reason the ADI is relatively slow is that it is necessary to integrate through several cycles of the heat to ensure steady-periodic condition is enforced. 

Equally important, the Green's function method is easy to parallelize while the ADI method is not. The ADI method is sequential by nature. It requires updating the entire grid to advance the time by $\Delta t$. In contrast, the Green's function involves a summation of the temperature contribution from each heated pixel, so the calculations for pixels are independent and easy to run in parallel. This parallelization is relatively simple to implement using the shared-memory multiprocessing library OpenMP.

In addition, we can parallelize other steps of the model. Propagation of the signal field array can be performed independently for each time slice of the field. Similarly, the laser gain calculation, modal content decomposition, and effective area calculation can be performed independently for each of the time slices.

\subsection{Execution speed considerations}

We have both MATLAB and Fortran versions of our model. Both versions use the FFT library FFTW\cite{FFTW}. Our experience is that FFTW is substantially faster than other FFT routines in Fortran.

OpenMP, the shared-memory multiprocessing library and compiler directives, is implemented in the version of Fortran we use, GNU Fortran 4.6.3. Using a desktop computer based on an Intel Core-i7 3770 (Ivy-Bridge) processor with four physical cores, we are able to obtain an excellent speed up. Using four threads instead of one, the time required to run the same fiber setup decreases by a factor of $\approx 3.2$.

Another important speedup was obtained by solving the thermal problem and laser gain equations on a coarser $z$-grid than the FFT propagation problem. This can dramatically reduce the model run time as well. Care in choosing this grid is important because a grid that is too coarse can reduce the computed gain and increase the instability threshold power.

The run times on our computer lie in the range of 0.25-1.5 hour/m depending mostly on the $z$ step size and the number of harmonics included in the Green's function. Larger cores use larger $z$ steps, and near-threshold runs require only a single harmonic, permitting times near 0.25 hour/m.

\section{Approximations of model}

All numerical models make judicious approximations. A brief list of ours follows:

Single $\lambda$ pump with single absorption and emission cross sections

Pump power is uniformly distributed across the pump cladding

All signal light is identically polarized

Steady-periodic heating is required for application of Green's function

Fixed period $\Upsilon$, which allows only signal frequency offsets $1/\Upsilon \times (0,\pm 1,\pm 2,...)$

Thermal boundary condition is fixed temperature on square boundary

Thermal boundary size approximately three times core diameter 

Heat equation solved in two dimensions (no longitudinal heat flow)

Thermal properties assumed uniform and isotropic

Temperature dependence of cross-sections, ${\rm d}n/{\rm d}T$, and $\tau$ not included

No refractive index dependence on $n_u$

Signal bandwidth $< 100$ GHz

Low contrast refractive index profile, \emph{e.g.} no air-holes as in PCF

Upper state population $n_u$ follows $I_s$ instantaneously (steady-state expression)

\section{Attributes of model}
Attributes of our model include:

Highly numeric - general and simple to add additional physical effects

Model a variety of refractive index, linear absorption, and doping profiles

Steady-periodic eliminates long integration times before steady state

Steady-periodic model produces well defined thresholds

The ADI method can be used to study transient behavior if desired

All transverse modes are automatically included

Thermal lensing automatically included

Comparatively short run times and minimal memory requirements

Variety of thermal boundary conditions possible using Green's functions

Green's function method offers large speedup using multiple processors

\section*{Acknowledgments}
This work partially supported under funding from the Air Force Research Laboratory Directed Energy Directorate.
\end{document}